\documentclass{aastex}
\newcommand{\beq}{\begin{equation}}
\newcommand{\eeq}{\end{equation}}

\newcommand{\grad}{\nabla}
\newcommand{\zhat}{{\bf \hat z}}

\newcommand{\xhat}{{\bf \hat x}}

\newcommand{\beal}{\begin{mathletters}\begin{eqnarray}}
\newcommand{\eeal}{\end{eqnarray}\end{mathletters}}
\newcommand{\bea}{\begin{eqnarray}}
\newcommand{\eea}{\end{eqnarray}}

\begin{document}
\title{Local hydrodynamics and operator stability of Keplerian
	accretion disks}
\author{Peter T. Williams\altaffilmark{1,2}}
\affil{Department of Astronomy and Department of Physics,
 University of Texas, Austin, TX  78712-1081, petwil@astro.as.utexas.edu}
\altaffiltext{1}{Now at Department of Physics, University of Florida,
	P.O. Box 118440,
	Gainesville, FL 32611-8440}
\altaffiltext{2}{WARNING!!!! This is an old (2000) manuscript, posted here for
archival purposes.  It is reproduced verbatim save for this footnote and the postscript which follows
the paper. Current address of author: 736 Spokane Ave, Albany, CA 94706 (no current affiliation)}
\shorttitle{Operator stability of Keplerian disks}
\shortauthors{Williams}
\begin{abstract}
We discuss non-self-gravitating hydrodynamic disks in the thin disk limit.
These systems are stable according to the Rayleigh criterion, and yet there
is some evidence that the dissipative and transport processes in these
disks are hydrodynamic in nature at least some of the time.
We draw on recent work on the hydrodynamics
of laboratory shear flows. Such flows are often experimentally
unstable even in the absence of a linear instability.
%According to recent theoretical work, t
The  transition to turbulence
in these systems, as well as the
large linear transient amplification of initial disturbances,
may depend upon the non-self-adjoint nature of the
differential operator that describes the dynamics of
perturbations to the background state.
We find that small initial perturbations can 
 produce large growth in accretion disks in the shearing
sheet approximation with shearing box boundary
conditions, despite the
absence of any linear instability. Furthermore, the differential
operator that propagates initial conditions forward in time is asymptotically
close (as a function of Reynolds number)
 to possessing growing eigenmodes. The similarity to the
dynamics of laboratory shear flows is suggestive that 
accretion disks might be hydrodynamically unstable 
despite the lack of any known instability mechanism.
\end{abstract}

\section{Introduction}
In this paper we discuss the hydrodynamic stability of thin, non-self-gravitating,
Keplerian accretion disks. The study of potential hydrodynamic and 
magnetohydrodynamic (MHD) 
instabilities in Keplerian shear flow is important because they may
explain transport within accretion disks.
(For a review of accretion disk physics,
see \citet{FKR:92}. A more recent review of accretion disk
stability and transport may be found in \citet{BaHa:98}.)
%Recently, researchers in the fluid dynamics and applied mathematics communities
%have developed novel theories of the transition to turbulence in shear
%flows \citep{TTRS:92}.
%We apply some of these new concepts to the old problem of
%the stability of accretion disks. 

The effective kinematic viscosity $\nu$ of accretion disks is traditionally
parametrized by $\alpha$ in the phenomenological relation
\beq
\nu=\alpha c_s H,
\eeq 
where $c_s$ is the sound speed and $H$ is the local disk scale height
\citep{ShSu:73}. Inferred values of $\alpha$ for dwarf nova (DN) accretion
disks, which are relatively well-studied and for which $\alpha$ is perhaps
the most observationally constrained, lie in the range of $\alpha \approx
0.1$---$0.001$. (See \citet{War:95} for an excellent review
of DN disks.) 
%The precise value depends upon
%whether the system is in outburst, among other things. Regardless,
These values for $\alpha$ are many orders of magnitude (e.g., 10--12)
larger than back-of-the-envelope estimates for the molecular
viscosity.
%, hence $\alpha$ is sometimes referred to as the anomalous viscosity.
Other disk systems,
such as the accretion disks in active galactic nuclei (AGN) and the
circumstellar disks in young stellar objects (YSO), are also thought
to have effective viscosities that are anomalously large in
the same sense \citep{Kro:99, Har:98}. 

It was originally hoped that the large Reynolds numbers in Keplerian
disks would be a source of hydrodynamic turbulence
%, leading to an anomalously large effective viscosity 
\citep{vWe:48}. However, 
%as has been repeatedly emphasized in the literature, 
Keplerian disks satisfy the Rayleigh
stability criterion for inviscid axisymmetric perturbations \citep{Cha:61},
\beq
{{d}\over{dr}}(r^2\Omega)^2 >0.
\eeq
It should be noted that this is a necessary, but not sufficient, criterion for
stability of the more general case of non-axisymmetric perturbations to
a viscous flow. Nonetheless, there is no well-established instability mechanism
in Keplerian hydrodynamic shear flow.

The canonical route to turbulence in disks
%, and the source of the anomalous viscosity $\alpha$,
is the Balbus-Hawley magnetic shearing instability \citep{BaHa:91, HaBa:91}.
However,
there may be some disks that are not sufficiently ionized for the Balbus-Hawley
instability to be effective. Ion-neutral collisions in circumstellar disks
associated with YSOs may dissipate the magnetic field too quickly for the instability
to take hold \citep{SGBH:00}. %In addition,
\citet{GaMe:98} simulated the dynamics of the prototypical DN system
\objectname{SS Cyg}, using a code developed by \citet{HMDLH:98}; based on their simulations,
they conclude that the Balbus-Hawley instability
may not be operating when the disk is in quiescence. 
In their disk simulations,
the magnetic Reynolds number drops below $10^4$, and
numerical MHD simulations \citep{HGB:96} indicate that the
Balbus-Hawley mechanism requires a magnetic Reynolds number $Re_B$ based
upon scale height to be larger than approximately $10^4$.
%Menou suggested, based upon
%the previous suggestions of (Spruit, ...), that a global hydrodynamic
%transport mechanism might be operating,
%namely spiral shocks or hydraulic jumps of some form. There is some
%evidence for this, as discussed by Menou, and as evinced in Doppler
%tomograms and eclipse maps of DN, although the evidence is not
%incontrovertible.

%We focus in this paper not on global hydrodynamic process, but on
%local ones. 
We show that the shearing-sheet approximation
with shearing-box boundary conditions shares
properties in common with such 
laboratory flows as Couette flow, Poiseuille flow, Hagen-Poisseuille
flow, and Taylor-Couette flow. (For 
%a discussion of these flows and their
%instabilities, see %any advanced textbook on hydrodynamic stability theory, such as 
an introduction to these flows, see
\citet{DrRe:81}.) These shear-dominated laboratory flows
exhibit transitions to turbulence even in the absence of a linear 
instability \citep{TTRS:92, RiZa:99}.
% as discussed by \citet{TTRS:92}, among others.
%Plane Couette flow  does
%not posess a linear instability, but has been observed to become turbulent
%at Reynolds numbers based upon channel half-width as low as 350
%\citep{LuJo:91}, as observed by \citet{TTRS:92}. Plane Poiseuille
%flow becomes
%unstable, according to classic eigenmode analysis, at a Reynolds number
%of 5772 \citep{Ors:71}, but may be seen to be turbulent in the laboratory at
%Reynolds numbers running from 1000 to 8000, approximately (see
%\citet{DaWh:28}, \citet{NII:75}, and other references in \citet{TTRS:92}).
Hagen-Poiseuille flow (pipe flow), for example,
 does not possess a linear instability, 
but is seen to become turbulent in the laboratory at a range of Reynolds
numbers, % with lower Reynolds numbers corresponding to pipes with rougher
%interior surfaces and with less-carefully conditioned flow at the pipe
%inlet,
 as observed by Reynolds \citep{Rey:1883}.% (see also the introduction
%to \citet{DrRe:81}).
%, the transition to turbulence in pipe flow is
%intermittent, the turbulence often occurring in bursts separated spatially
%by laminar sections of flow. At higher Reynolds number the bursts are
%more frequent and occupy a larger fraction of the flow, until the flow
%finally becomes fully turbulent.
% Taylor-Couette flow (flow between two rotating
%cylinders) is seen to be turbulent in some instances at high Reynolds
%number in regions of the parameter space that are stable according
%to the Rayleigh criterion \citep{Col:65}.
%This transition is seen to exhibit a hysteresis effect also, so that
%the presence of turbulence is determined not just by the control parameters
%of the flow but also by the history of the system (H. Swinney, personal
%communication).
%These behaviors suggest that the transition to turbulence in these
%flows does not necessarily occur at a single critical Reynolds number through
%a linear instability. 

One property of these flows
\citep{Far:88, TTRS:92, GeGr:93, RSH:93, BrGr:99, Gro:2000}
% is that the linear operator
%governing the growth of perturbations to the background flow is not a
%normal operator in the formal mathematical sense. The eigenmodes of
%these operators are highly collinear, so that eigenmode analysis is 
%a poor technique for exploring the dynamics of perturbations. 
is that initial
perturbations may undergo large transient growth in the absence of
any growing eigenmode. Furthermore, the eigenvalue spectrum of these
operators is very sensitive to perturbations to the operators,
so that the operators are in many cases asymptotically close (as a function
of Reynolds numbers) to operators that possess growing eigenmodes, even
if the operators themselves do not have growing modes.

%We apply some of these recently developed concepts regarding the stability of shear
%flows to the problem of Keplerian shear flow. We specifically consider the behavior
%of 2D ($x$--$y$) perturbations in the shearing-box system, which has
%been used extensively in simulations of local accretion disk hydrodynamics
%and MHD, and we ask whether the operator governing such perturbations is
%stable (in the spectral sense) to small perturbations to that operator.

\section{Local hydrodynamics and shearing box}
%We consider the non-self-gravitating, thin-disk approximation, and
%assume incompressibility. We use the
%shearing sheet approximation. This
The shearing-sheet approximation
 replaces the $(r,\theta)$ coordinates
in the disk with a local Cartesian plane $(x,y)$
(see e.g. page 439 of \citet{RyGo:92} and references therein). 
%It is the first-order
%approximation to the disk dynamics in the parameter $\ell /r$, where
%the $\ell$ is the maximum lengthscale in $x$ and $y$ for perturbations
%of the flow variables.
We neglect the $z$ dimension and
assume incompressibility. We have then
\beq
\partial_t {\bf v} + {\bf v} \cdot \grad {\bf v} + \rho^{-1}\grad p +
2\Omega\zhat {\bf \times} {\bf v} - 4 A\Omega x \xhat = \nu \grad^2 {\bf v},
\label{eq:shsh}
\eeq
where $A$ is Oort's first galactic constant,
\beq
A \equiv -\textstyle{{1}\over{2}}r\partial_r \Omega \Big\vert_{r_0}
= {{3}\over{4}}\Omega_0.
\eeq
The background flow is $-2Ax{\hat {\bf y}}$. Perturbations ${\bf u}$ to this flow
may be described in terms of a streamfunction $\psi$, so that ${\bf v} =
-2Ax{\hat {\bf y}} + {\bf u}$, and $(u_x,u_y)=(\partial_y\psi,-\partial_x\psi)$.
We let
\beq
\psi(x,y,t)=\phi(x,t)e^{i\beta y},
\eeq
and we non-dimensionalize by measuring time in units of
$1/\Omega$ and distance in units of some unspecified $L$, which we may
take to represent, say, the disk scale height.
 The dimensionful viscosity $\nu$ is
the inverse Reynolds number $\nu \rightarrow R^{-1}$
with the length $L$ as the defining lengthscale. In the case that
$L$ is the disk scale height, then $R$ may typically lie in
the range $10^{10}$--$10^{12}$ approximately, as discussed above.

Taking the curl of (\ref{eq:shsh}) and neglecting $\partial_z u_z$
in the spirit of a 2D analysis, we have then that
\bea
 \partial_t \phi = 
\left[\partial_x^2 - \beta^2 \right]^{-1}
 i \beta \left[ (i \beta R )^{-1} (\partial_x^2 - \beta^2)^2 - 
2{A} x (\partial_x^2 - \beta^2) \right] \phi,
\label{eq:OS}
\eea
which is a form of the Orr-Sommerfeld equation for viscous
shear flow (Reddy et al. 1993).
 (Note that in Keplerian disks, the nondimensionalized
$A$ is simply ${A} = 3/4$.) If we further separate variables,
this equation is seen not to be formally self-adjoint in $x$ in
the $L_2$ norm for $\phi$. With the addition of rigid boundaries
in $x$, this equation governs 2D perturbations to Couette flow;
at high Reynolds number, its eigenmodes are highly collinear, and
its eigenvalues are extremely sensitive to perturbations to the
equation.

We seek solutions to equation (\ref{eq:OS}) in the shearing-box system.
This is the shearing sheet with the addition of periodic boundary
conditions in $y$ and sheared periodic boundary conditions in $x$ \citep{BaHa:98}.
Specifically, we consider a square domain of size $L$, and the
boundary condition in $x$ is 
\beq
\psi(0,y,t) = \psi(L,y-2ALt,t).
\eeq
Equation (\ref{eq:OS}) is not separable in the shearing box
system. However, a complete set of solutions exist; these
are the Kelvin modes \citep{Kel:1887, FaIo:93}. These modes
 are Fourier modes in $x$ and $y$, where
\beq
\psi(x,y,t) = C_{k(0),\beta}(t) e^{ik(t)x + i\beta y},
\eeq
with $k(t) = k(0) + 2 A \beta t$,
and
\beq
C_{k(0),\beta}(t) = C_{k(0),\beta}(0) {{k(0)^2+\beta^2}\over
{k(t)^2 + \beta^2}} \exp\left[-R^{-1} \int_0^t \left(
k(\tau)^2 + \beta^2 \right) \ d\tau \right].
\label{eq:kelvin}
\eeq
To satisfy boundary conditions, we have $\beta = m 2\pi$
and $k(0) = n 2\pi$. At times $t=j\Delta t$, where $j$
is an integer and $\Delta t = (2Am)^{-1}$, the streamfunction may be written
\beq
\psi(x,y,t)=c_{n,m}^{[j]} e^{2\pi i(nx+my)}.
\eeq
We restrict our attention to the invariant
subspace $m=1$, since these are the least dissipated modes with
the potential for growth, and we henceforth drop the subscript $m$. 
The streamfunction at timestep $j$ is then completely
specified in terms of the state vector ${c}^{[j]}=
(\ldots,c^{[j]}_{-1},
c^{[j]}_{0},c^{[j]}_{1},\ldots)$.
The solution to equation (\ref{eq:OS})
may now be written as a difference equation in $t$,
\beq
%{c}\left((j+1)\Delta t\right)=
{c}^{[j+1]} = {\cal L}{c}^{[j]},
\eeq
where ${\cal L}$ is a matrix operator.
It is zero everywhere except on the subdiagonal,
reflecting the fact that the $i^{\rm th}$ element of the
state vector ${c}^{[j]}$ is taken from the $(i-1)^{\rm th}$ element
of ${c}^{[j-1]}$ (i.e. ${c}$ on the previous timestep),
 adjusted in amplitude according
to equation (\ref{eq:kelvin}). The matrix ${\cal L}$ provides a convenient representation
of the propagator for the initial value problem:
\beq
%{c}(n \Delta t) = {\cal L}^n {c}(0).
{c}^{[j]} = {\cal L}^j {c}^{[0]}
\label{eq:propagator}
\eeq
The operator ${\cal L}$ is formally generated by the Orr-Sommerfeld operator
${\cal L}_{\rm OS}$ (which is the right-hand side of 
equation {\ref{eq:OS}}),
in the subspace of discrete Kelvin modes described above:
\beq
{\cal L} = e^{\Delta t {\cal L_{\rm OS}}}.
\eeq
The Laplacian in the operator ${\cal L_{\rm OS}}$
is rendered well-defined by the boundary conditions.

\section{Transient growth and operator perturbations}
In order to discuss growth, we first need a norm to measure the
strength of perturbations. We use the $L_2$ norm of the streamfunction
$\psi$. In this norm, the magnitude of a perturbation ${c}$ is
\beq
\|{c}\| = \int \int \psi^* \psi\ dx\ dy = \sum_{i=-\infty}^{\infty}
c_i^*c_i.
\eeq
Given this vector norm, the usual operator norm of a 
matrix ${\cal A}$ is defined using the vector norm:
\beq
\|{\cal A}\| = {\rm max}_{\|x\|=1} \|{\cal A}x\|, 
\eeq
\notetoeditor{The output for equation 15 is not quite what I would
like: The ``max'' needs to be directly over the ``$\|x\|=1$'',
but I don't know how to do that.}
and it is equal to the maximum singular value of ${\cal A}$.
The norm $\|{\cal L}^j\|$ measures the maximum possible growth
of perturbations in $j$ timesteps.

In figure 1, we plot (triangles) the operator norm
$\|{\cal L}^j\|$ versus $j$, for a Reynolds number of $10^6$.
The ``hump'' is a typical characteristic behavior for
the norm for a non-normal matrix.
%In general, given an equation of the form of equation 
%(\ref{eq:propagator}),
%we might have suspected that growth can only occur in the
%case that ${\cal L}$ has eigenvalues with magnitude larger than
%one. This intuition is based upon the behavior of normal operators;
%it does not hold in the case of non-normal operators.
%
%The amount of transient growth of initial perturbations can be
%quite large.
We show in figure 2 (triangles) the log of the amount of
growth that can be obtained as a function of Reynolds number,
which we denote by $G(R)$, where $G(R)={\rm max}\ \|{\cal L}^j\|$
and the maximum is taken over $j$.
As ${\cal L}$ is represented by a matrix that is zero everywhere
except on the subdiagonal, powers of it are easy to compute.
%,
%and we are able to do so while keeping track of on the order
%of $10^4$ modes, which guarantees convergence over the range
%of Reynolds numbers shown.
The maximum growth scales as
\beq
G(R) \sim R^{2/3},
\eeq
where the maximum is taken over the values of $j$. This result
is also easily obtained analytically from equation (\ref{eq:kelvin}) in
the limit of large $R$. Specifically, for large $R$, we have
\beq
G(R) \approx (0.092) R^{2/3}
\eeq
 The line drawn through the triangles in
figure 2 is the analytic approximation above. 

This type of transient growth has been remarked upon previously
in the literature; it is merely a consequence of the kinematics
of the Kelvin modes as described by equation (\ref{eq:kelvin}).
According to the Rayleigh hypothesis, instability occurs whenever
there exists an unstable mode. 
%Realistically, all modes are expected
%to be excited with some nonzero amplitude, although that amplitude
%may be quite small. Exponential growth simply implies that at some
%finite time, higher-order effects must become important, leading
% to a secondary flow as in the case of exchange of stabilities,
%additional higher-order instabilities, or turbulence.
The unbounded
exponential growth of an unstable mode
is of course unphysical.
%; it simply guarantees that
%the initial perturbations may be vanishingly small.
Large transient
growth in the linear limit, as obtained here, simply implies that
the strength of the initial perturbation required in order to reach
the nonlinear regime is potentially very small, but not infinitesimal.
Nonlinear effects may then positively or negatively feedback upon
the original perturbation, but we do not discuss these effects here.

%What has perhaps not been appreciated to such a great extent is the
%implication of t
This transient growth has implications for the spectra of
 perturbations to the
operator ${\cal L}$.
We construct matrices ${\cal M}$ such that
each element of ${\cal M}$ is randomly drawn from the unit disk in the
complex plane. We then construct the perturbed operator $\tilde{\cal L}$,
\beq
\tilde{\cal L} = {\cal L} + \epsilon {\cal M}.
\eeq
The matrix ${\cal M}$ couples the Kelvin modes, allowing the possibility
for feedback to the transient growth observed in $\|{\cal L}^n\|$.
Note that we do not rescale ${\cal M}$ by its operator norm, 
$\|{\cal M}\| \approx \sqrt{2\ {\rm dim}({\cal M})}$.

Although in principle ${\cal M}$ need not have any
eigenvalues with magnitude larger than one, in practice it always will. Given this, for each
${\cal M}$ there exits a positive $\epsilon$ such that the operator
$\tilde{\cal L}$ possesses an eigenvalue $\lambda$ with $|\lambda| > 1$,
so that the operator $\tilde{\cal L}$ possesses a growing
eigenmode. Such an eigenmode corresponds to a growing Bloch mode,
\beq
\phi(x,t) = \xi(x,t) e^{i \omega t},
\eeq
where $\xi(x,t)$ is periodic in $t$ with period $\Delta t$.
In figure 1 (open squares), we show the growth of the operator
norm $||({\cal L}+\epsilon{\cal M})^n||$ at a Reynolds number
of $10^6$ and with $\epsilon = 5 \times 10^{-4}$. The behavior
of this perturbed operator diverges exponentially from that of the
unperturbed operator in the operator norm, at large times. This
reflects the presence of a growing eigenmode.
Interestingly, the minimum values of $\epsilon$ (denoted
$\epsilon_{\rm crit}$) required
to produce a growing eigenmode for the operator $\tilde{\cal L}$
do not depend greatly upon the matrix ${\cal M}$, for our choice
of probability measure for ${\cal M}$. Furthermore, the 
values $\epsilon_{\rm crit}$ become asymptotically small as the Reynolds
number is increased.

In figure 2 (circles) we plot the log of $1/{\epsilon_{\rm crit}}$
versus the log of the Reynolds number. We perform this calculation
for forty different random matrices for each Reynolds number.
(This is a more computationally
intensive operation than the determination of the
maximum operator norm ${\rm max}\|{\cal L}^j\|$. We have made
extensive use of the LAPACK routine package
\citep{lapack:99}, running on the beowulf cluster parallel virtual
machine in the University of Texas Astronomy Department.) 
The points are the median values of $1/{\epsilon_{\rm crit}}$,
and the error bars are the first and third quartile.
The 
relationship is
 a power law with an index of $0.88 \pm 0.01$. Deviations from this
power law at the low end reflect the effects of the
discretization of time into timesteps $\Delta t$.
%, as Kelvin modes
%at these low Reynolds numbers grow and die in times less than
%$\Delta t$. 
Deviations from the power law at high Reynolds 
number are due to computational limits on the necessarily finite
size of the matrix representations for ${\cal L}$ and ${\cal M}$,
as we have confirmed by varying these dimensions. The corresponding
points for larger or smaller matrices fall on top of the points shown
here, up until the point at which they turn off of the power law
relationship. This point is determined by the value for $j=j_{\rm peak}$ at
which the operator norm $\|{\cal L}^j\|$ reaches its maximum, as matrices
of order smaller than $j$ will not faithfully represent the growth
of $\|{\cal L}^j\|$. In figure 2 we also show the dependence
of $j_{\rm peak}$ upon Reynolds number (squares);
 it is seen to depend upon
$R$ in the form $n \sim R^{1/3}$, and the line is the theoretical
result for large Reynolds number. 
%The fact that this exponent 
%equals the difference between the exponent in the power law for
%$1/\epsilon$ and the power law for ${\rm max} \|{\cal L}^j\|$
%is not a coincidence.
Given maximum growth $G(R)$ of the operator
norm for the unperturbed operator, we need only perturb the operator
by a matrix that is zero everywhere except at the matrix element that
couples Kelvin modes at the end of their growth period to the modes
that are just beginning their growth period , i.e. $j_{\rm peak}$
timesteps earlier. This element need only have an amplitude of
$\epsilon > 1/G(R)$,
to create a feedback loop.
Rougly, the number of matrix elements that will effectively
couple grown modes to the earlier maximally amplified modes may
be expected to scale as $j_{\rm peak}^2$; when added in quadrature,
the effective feedback then would scale as $j_{\rm peak}G(R)$, which
grows as $R^{1.0}$. However, each of these matrix elements do not have the
same potential for producing feedback to the Kelvin modes, so that
the actual scaling of $G(R)$ with $R$ is somewhat less steep.

One physical interpretation of the operator perturbation $\epsilon
{\cal M}$ is that it potentially represents the differences between the
idealized operator ${\cal L}$ and the true operator that governs
the system. 
Compressibility, tides, magnetic fields, and so forth, will affect
the operator ${\cal L}$. While we cannot say that any of these
effects will actually result in a new operator $\tilde{\cal L}$ 
that possesses growing modes, we can say that, at least in the
probability measure of random matrices ${\cal M}$ that we have
discussed, most perturbed operators of the form $\tilde{\cal L}
= {\cal L} + \epsilon {\cal M}$ have growing eigenmodes for
asymptotically small $\epsilon$ at high Reynolds number. 
%(We also note that compressibility is known to destabilize
%Couette flow \citep{HuZh:98})
This
result reflects properties of ${\cal L}$ itself. Had ${\cal L}$
been a normal operator, this result would not obtain, and the
spectrum of ${\cal L}$ would be very stable to operator perturbations
of the form described above.

%We hope to extend this work by including the full space of perturbations,
%rather than a restricted subspace. It would be particularly nice to
%solve the 3D problem, which would include the Taylor modes. 
%We would also like to extend our consideration to other norms besides
%the $L_2$ norm. Of particular interest is the energy norm, which
%measures the growth of the kinetic energy in the perturbation.

\section{Conclusions and Discussion}
The shearing sheet system with
shearing box boundary conditions appears, as hypothesized, to share
many properties in common with such flows as Couette flow
and other shear-dominated flows.
% The linearized perturbation operators in these
%flows are non-normal; in the nominally stable regions of
%parameter space, the systems are capable of large transient
%growth in the $L_2$ operator norm and the operators are asymptotically
%close to operators possessing growing eigenmodes.
The operator 
${\cal L}$ discussed above is clearly not normal. It does not even
possess eigenmodes. However, as we have shown, ${\cal L}$ is
capable of large transient growth in the operator norm, and it is
within $\epsilon$ of possessing growing eigenmodes in the sense
described above, where $\epsilon \sim R^{-1}$. In particular,
for the range of Reynolds numbers expected in disks as
discussed above, $\epsilon$
is on the order of $10^{-8.5}$.

This suggests that Keplerian shear flow may share the same phenomenology
as the above-mentioned classic laboratory flows, including turbulence
in the nominally stable regions of parameter space. 
%As previously
%mentioned, other effects observed in these flows include the 
%presence of intermittent turbulent bursts and hysteresis effects.
We hypothesize that non-magnetized accretion disks may therefore
be turbulent in the absence of a linear instability, as suggested
previously by \citet{Dub:92} and \citet{RiZa:99}. This could
explain the presence of transport in DN in quiescence, as well
as transport in other cool disks such as those associated
with young stars.
We suggest that simulations of 2D vertically-symmetric hydrodynamics in the
shearing box system should exhibit large transient growth of initial
perturbations and operator sensitivity as described above.

%%%%%%%%%%%%%%%%%%%%%%%%%%%%%%%%%%%%%%%%%%%%%%%%%%%%%%%%%%%%%%%%%%%%%%%
\acknowledgments
%%%%%%%%%%%%%%%%%%%%%%%%%%%%%%%%%%%%%%%%%%%%%%%%%%%%%%%%%%%%%%%%%%%%%%%
{\it
The author wishes to thank the Theory Group of the Department of
Astronomy at the University of Texas for their continued support,
and for their generosity with the group computing facilities.
He also wishes to thank Ralph Showalter of the Math Department for
many helpful discussions.
}

\figcaption[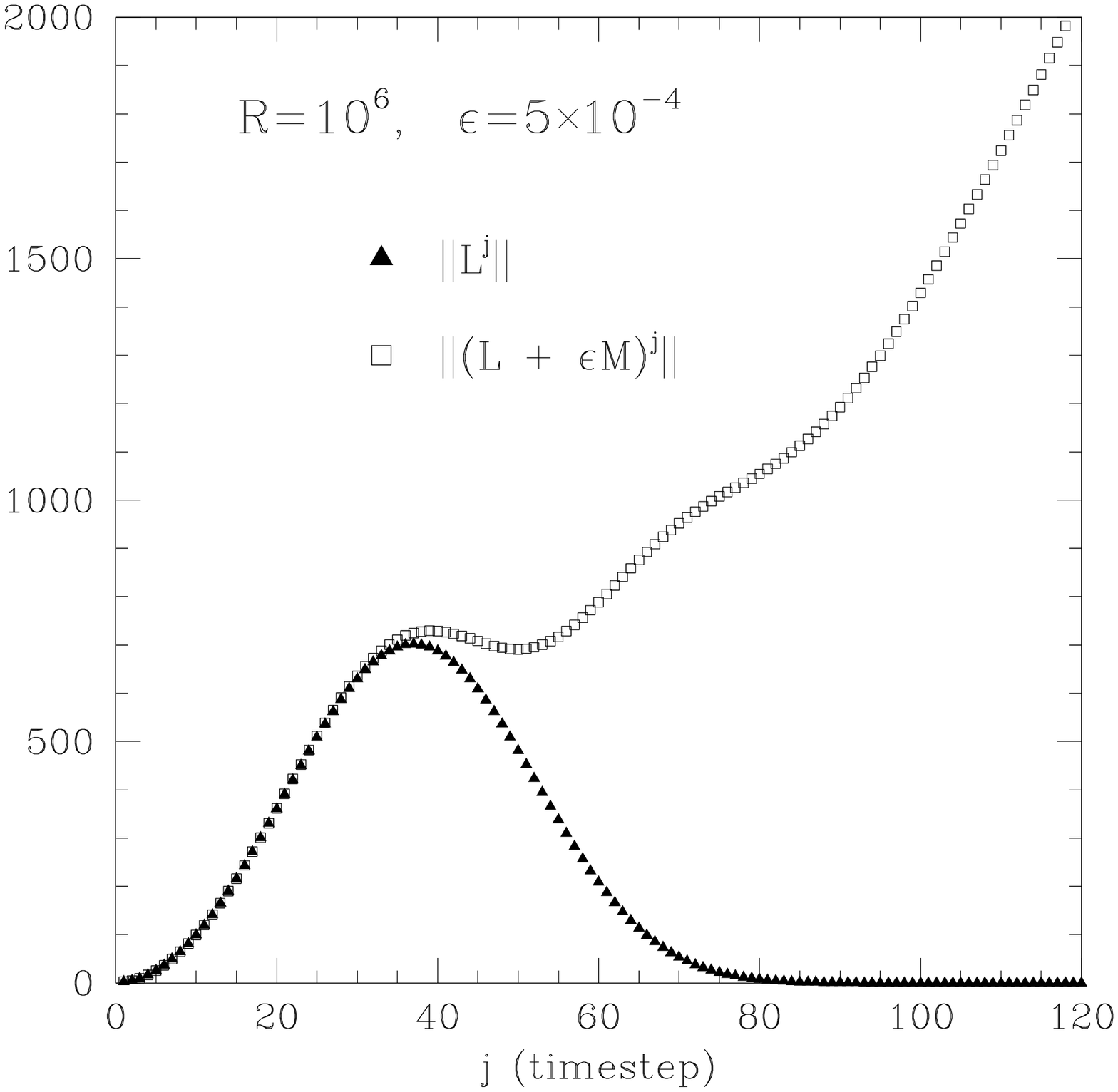]{$\|{\cal L}^j\|$ vs. $j$ for $R=10^6$. Triangles are the
unperturbed operator, squares are for the operator perturbed by
a random matrix with matrix elements taken from the unit disk
and scaled by $\epsilon = 5 \times 10^{-4}$.
Resolution is 201 modes.\label{fig:amp}}

\begin{figure}
\figurenum{1}
\epsscale{0.5}
\plotone{amp.eps}
\caption{}
\end{figure}

\figcaption[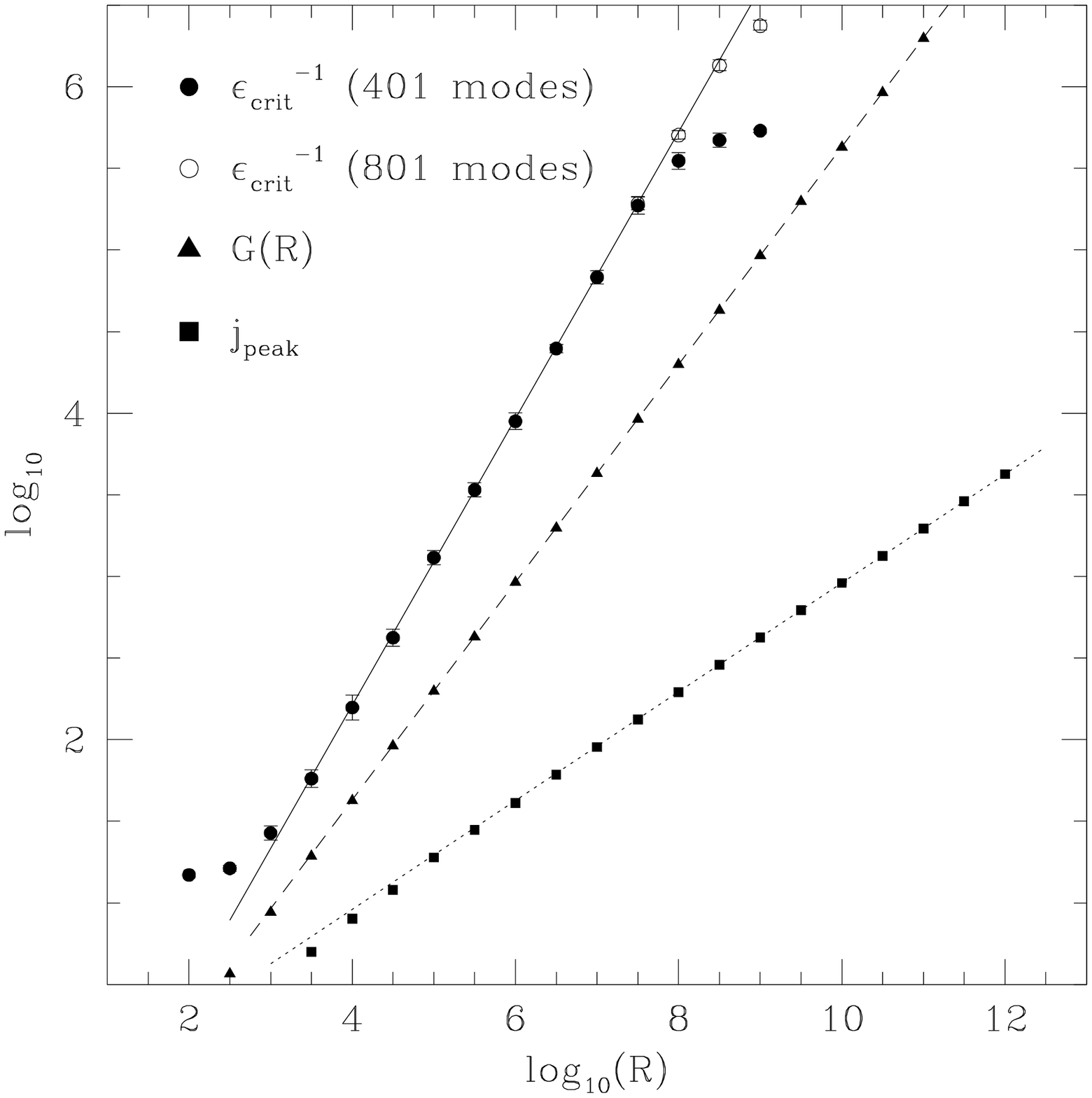]{Log-log plot of 
$\epsilon_{\rm crit}^{-1}$
(closed and open circles), $G(R)$
(triangles), and $j_{\rm peak}$ (squares) versus Reynolds number $R$.
Dashed line and dotted
line are theoretical curves for large $R$ limit for
$G(R)$ and $j_{\rm peak}$ respectively,
and solid line is power law fit to the central part
of curve for $\epsilon_{\rm crit}^{-1}$. Deviation from power law behavior
at high $R$ is due to the limitations of numerical resolution,
as discussed in text. Closed and open circles are results from runs with
401 and 801 modes respectively. \label{fig:svd}}

\begin{figure}
\figurenum{2}
\epsscale{0.5}
\plotone{svd3.eps}
\caption{}
\end{figure}

%\begin{figure}
%\figurenum{3}
%\epsscale{1.}
%\plotone{ran.eps}
%\caption{caption here\label{fig:ran}}
%\end{figure}

\newpage
\section{POSTSCRIPT}
The document appearing on the previous pages was
submitted to ApJL on 19 Oct 2000 and appears here now without alteration.
This paper was not accepted for publication.
The primary justifications for this decision were that
(1) the streamfunction norm is not a physical norm, (2) the measure of perturbations
in the form of random matrices as described does not have an obvious physical basis,
(3) it was suggested that the 2D perturbations presented here are essentially
uninteresting and that the perturbations should be performed in 3D. Note, for
example, that the perturbations that experience the greatest transient amplification
in planar Couette flow consist of shear-aligned vortices. These are inherently three-dimensional
structures, Squire's theorem not being relevant. However, 3D streamwise vortices in Keplerian shear
flow are quite different from similar perturbations to plane Couette flow, because of the
presence of an epicyclic frequency, so to an extent this is where the interesting physics is.

It was also pointed out that the bootstrapping described here ``cannot be regarded
as a proof of nonlinear instability.'' On that the referee and I are wholly agreed. If I implied 
in my paper that I thought differently, it was unintentional.

In  any case, the first objection listed above is quite valid, but on the other hand this problem is easy enough to fix,
and switching to a more physical norm such as the energy norm has little material effect on the results.

The second objection is also valid. The problem is that, as the referee put it,
``a random perturbation of a system of equations embodying a conservation law will allow an
artificial growth of the previously conserved quantity.'' True. (Although, strictly
speaking, here the unperturbed system of equations is dissipative.)
This defect might be remedied by restating the perturbations in terms of,
say, noise in the form of random forcing in the system.
Putting the perturbations on more physical grounds would have made the
paper much better, and probably would not have required too much work in the end.

The third objection is quite obviously more substantial in the obstacles it presented to me.
When I submitted this paper, I had already accepted employment elsewhere working in a completely different subject area.
I felt I could fix the other problems raised by the referee, but
a 3D analysis would take {\em substantially} more work;
without question I did not have the time to do such an analysis in place of the 2D results
presented here.

I was at the time completely naive about the publication process. 
This was the first paper I had ever submitted, as is evident by my lack of familiarity
with the idiom. Although I was very upset by the report,
I simply deferred to the referee on this matter. I reluctantly abandoned the paper.
Perhaps he was right, I thought, and only a 3D analysis would be worthy of publication
in the peer-reviewed literature.

I sincerely wish to thank the referee for the time and effort he took for 
his quite thorough and penetrating analysis.
In reading the report over now,
I see that he was trying to make helpful and encouraging suggestions, but all I heard at the
time was a flat ``no.'' 
Understanding the editorial process, and
reading between the lines of referee reports, it appears, are acquired arts. 
On the referee's third objection above, however, I must now respectfully disagree.
With perhaps a fair bit of polish, following the remaining suggestions of the referee, 
this modest work might have been a helpful addition to the literature at the time.
I believe now, as I did in October of 2000, that
even a 2D analysis of this system was interesting and worthy of appearing in the peer-reviewed literature.

Apparently I am not alone.
In the intervening years, other researchers have performed similar, although admittedly more insightful and extensive,
analyses of transient amplification in this 2D system. I mention a few noteworthy papers below, some of which
were accepted for publication by the referees and editors of other journals.
My list of relevant papers is certainly not exhaustive and I apologize to anyone I have left out.
I have been quite happy to see these papers come out, because I believe in the value of this type of analysis.
It has been illuminating to see the different approaches that people have taken to this problem. I anticipate that this
line of work will ultimately lead to a greater understanding of transport in cool disks.

There are many stories behind my paper, each with its own amusing
ironies. These in turn mostly serve now as humorous little lessons to me that I will largely
keep to myself. As I have no
intention whatsoever of resuming the work presented here, I have decided simply to post it for
archival purposes. A more extended discussion appears in the second chapter of my dissertation.
So, if anybody wonders why I did not publish my dissertation, here at least is part of the
answer. :) 

My graduate research advisors, J. Scalo and J. C. Wheeler, deserve my ample thanks for their support. 
My work certainly did not benefit either of their research programs; I had conceived
of the project long before I ended up under their wings. In retrospect, it was clearly
a mistake on my part to continue with a project I had begun earlier in my career,
but ``it seemed like a good idea at the time,'' as they say.
 As I worked autonomously on this project, any
mistakes or other gaffes in this paper are entirely my own.

Thanks go to Matt Umurhan for pointing out to me several of the references
listed below, and for some fun and enlightening conversations on related topics.
I also thank the reader for enduring my effusive commentary. I felt I couldn't post
this paper without an explanation.

{\noindent
Peter Williams\\}
Albany, CA

%%%%%%%%%%%%%%%%%%%%%%%%%%%%%%%%%%%%%%%%%%%%%%%%%%%%%%%%%%%%%%%%%%%%%%%
%%%%%%%%%%%%%%%%%%%%%%%%%%%%%%%%%%%%%%%%%%%%%%%%%%%%%%%%%%%%%%%%%%%%%%%
\end{document}